# Real-Time Detection and Classification of Astronomical Transient Events: The State-of-the-Art

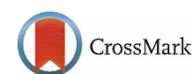

*Gianmario Broccia\**

*M.Sc. in Energy Engineering, Graduated at University of Cagliari, Cagliari 09123, Italy

A B S T R A C T

In the last years, the need for automated real-time detection and classification of astronomical transients began to be more impelling. Better technologies involve a higher number of detected candidates and an automated classification will allow dealing with this amount of data, every night. The desired state-of-the-art in detection and classification will be presented in its key features and different practical approaches will be introduced, as well. Several ongoing and future surveys will be presented, showing the current situation of Time-Domain Astronomy, and eventually compared with the desired state-of-the-art. The final purpose of this paper is to highlight the general technology readiness level with respect to the level yet to be achieved.

## 1. Introduction

The modern study of astronomical transient events started within the '800 although such kind of phenomena was already known, especially in the form of Supernovae. Among the most famous events in this sense are SN1572, observed by Tycho Brahe in the Cassiopeia Constellation, and SN1604, the last known Supernova exploded inside the Milky Way and studied by Johannes Kepler in 1604. With the birth of the telescope, owed to Galileo Galilei in 1609, the astronomers were finally able to better approach these events.

Improvements in optics and the space age led to an extraordinary increase in the number of detections per night, up to hundreds of transients in the current period and thanks to the constantly improving technology, the number of transients detected per night is estimated to rise to millions **(1).**

Astronomical transient events are one of the most interesting topics studied by Time-Domain Astronomy.

---

\* *Corresponding author.* Linkedin: https://www.linkedin.com/in/gianmario-broccia-827737165/
E-mail address: gianmario.broccia@gmail.com

To explore the subject, this article will show, in the first instance, the nature of astronomical transients by presenting the most representative classes. Secondly, the features that an optimal data reduction pipeline should have will be also presented, to give a reference for comparison with ongoing projects. Several surveys will be presented and characterized, especially about data processing: Catalina Real-Time Sky Survey, TESS, Antarctic Survey Telescopes, LSST, Pan-STARRS, Spitzer Space Telescope, Zwicky Transient Factory, Colorado Ultraviolet Transit Experiment, Sardinia Radio Telescope, AGILE, ROTSE, James Webb Space Telescope, Very Large Array, WFIRST and EUCLID.

Finally, in the "Discussion" paragraph, conclusions will be drawn regarding the readiness of the current and coming technology, in comparison with the desired state of the art. The surveys will be first presented in their features and only then a comparison will be made, calling into question each survey again. Any reference lacking DOI has been provided with a bibcode, ISBN code, or tagged as "Unpublished" either to indicate that its nature did not need any peer-review publication (ex: handbooks) or the final draft was available only on ArXiv.

## 2. Astronomical Transient Events

An astronomical transient event (hereafter "transient") is commonly defined as a natural phenomenon with a duration ranging from milliseconds up to months. Their detection is owed to the emission of either electromagnetic or gravitational waves, which makes them particularly bright (when visible in the optical band) and visible from cosmological distances.

In the following paragraphs, the most known classes of transients will be briefly introduced.

### 2.1 Novae, Kilonovae, Supernovae, Hypernovae

**Novae**. The definition of Nova is owed to Tycho Brahe who observed the Supernova SN1572 mentioned before, in 1572. The article derived from those observations was indeed named "De Nova Stella" **(2)**, to indicate a star that appeared apparently from nowhere. Ironically, the process that triggers a Nova originates in a binary system that involves a White Dwarf (WD), the remnant of a Sun-like star, and a Red Giant (RG), which is a Sun-like star in its final stages **(3)**.

The older WD collects material (mostly Hydrogen) from the companion, creating a layer of stolen gas on its surface. This process is usually allowed by the close distance between the stars, which implies an orbital period in the order of hours **(4)**. When the temperature reaches 20 million Kelvin, new fusion reactions take place, and ~3-33 $M_\oplus$ worth of material is expelled at thousands of kilometers per second out of the system **(3)**. The resulting increase in the magnitude makes the star appear like a new object and it can last several weeks. This happens in a cycle which is usually regular and characteristic of each system.

**Kilonovae.** A Kilonova is an event in which two supermassive objects merge. Such objects can usually be neutron stars or even black holes and the resulting luminosity is 1000 times a classical nova, but still no more than 1/10 of a supernova (**5**). After the event GW170817, which allowed LIGO and Virgo observatories to detect gravitational waves for the first time, Kilonovae became one of the suspected sources of Gamma-Ray Bursts (**6**).

**Supernovae.** A Supernova is commonly intended as the explosion of a massive star. Although a Supernova can belong to several classes, only the most known is presented: the Type II Supernova.

A star with a mass at least 9 times the mass of the Sun (**7**), arrived at the end of its life, is no longer able to burn, having converted all the hydrogen in helium. In this case, the gravitational force is no longer countered by fusion reactions and the nucleus begins to contract under its own weight, starting a cycle in which the contraction triggers new fusions of increasingly heavy elements. When the iron-56 forms, no new reaction can take place, and nothing can stop the final contraction of the ferrous nucleus of the star (**8**). If the nucleus overtakes the Chandrasekar Limit (1,44 Solar Masses), an intense explosion takes place and the surrounding material is jettisoned to relativistic speeds from the star, creating a shock wave and a Type II Supernova (**9**).

During this event, the dying star emits in an instant as much energy as it would emit the sun in all its entire life.

**Hypernovae.** Hypernovae may be generated by exceptionally massive stars (> 30 M☉) and the consequent explosion is expected to be 100 times more powerful than a regular Supernova. Eta Carinae in the Constellation Carina is the most massive known star, with a mass at least 100 times the mass of the Sun, and it is the strongest candidate to be a Hypernova in the near future (**10**).

*2.2 Tidal Disruption Events*

A Tidal Disruption Event (TDE) is a phenomenon that occurs when a star passes so close to a massive Black Hole to cross its Roche Limit, suffering from intense forces that eventually lead to its destruction (**11**).

In 1975 a first proposition suggested that any galaxy with a supermassive black hole at its center would be subject to tidal disruption events and the consequent flares from the stellar remnants would be a clear indicator (**12**).

*2.3 Transits*

In astronomy, a transit is commonly intended as the event in which a body of interest passes between a star and the observer (**13**). Inside our solar system, only the Moon, Venus, and Mercury are capable to create this phenomenon.

Outside the solar system, the transits are a valuable tool to verify the presence of exoplanets around a known star. By Transit Photometry it is possible to measure the dimming of the host star during the transit itself and obtain a value for the radius of the planet, while the mass can be obtained by Transit Spectroscopy (**14**). The density is then obtained and several hypotheses on the nature of the planet can be made.

Transits can also happen when two stars in a binary system occult each other or when a Jovian satellite produced an equivalent shadow zone on Jupiter. As a matter of example, along with the period 1985-1990, a series of occultations occurred between Pluto and its major satellite Charon allowed to precisely deduce the physical parameters of the two bodies for the first time **(15)**.

*2.4 Gamma-Ray Bursts*

The first detection ever of Gamma-Ray Bursts (GRBs) dates back to 1967 when the U.S. Vela satellites, purposed to monitor the use of Soviet nuclear weapons, discovered instead strange gamma beams from the deep space **(16)**.

These events last from milliseconds to minutes and are known to be the most energetic ever observed, although their origin is still not clear. Many theories were proposed, implying black holes evaporation, exotic types of supernovae, accretion of neutron stars, etc **(17)**.

The light curve of a GRB does not show any particular pattern and appears to be aleatory and under no circumstances, two GRB will produce a similar light curve **(18)**.

*2.5 Fast Radio Bursts*

Fast Radio Bursts (FRBs) extremely rare high-energy phenomena that manifest with transient radio impulses of a very short duration **(19)**.

The first detection happened in 2007 (FRB 010724) and, until now, only a fistful of new events was detected.

FRBs appear to be point source-like and are characterized by a wide range of radio frequencies, usually around 1400 MHz, though some were detected at frequencies in the range of 400-800 MHz **(20)**. They usually appear as a flash of energy that runs out in milliseconds without showing any change in intensity. Moreover, they do not appear to come from a specific region, being instead widespread all over the sky. As of 2019, no explanation is commonly accepted even though many hypotheses are being considered.

*2.6 Gravitational Microlensing*

Gravitational Lensing, initially foreseen by Einstein in 1912 and predicted by the general theory of relativity, is represented by the bending of a light beam in presence of a massive object (for example, a galaxy) between the target object and the observer. Such bending allows the observer to see an object that is covered by another one.

Microlensing is a sub-class of gravitational lensing and it occurs when a far less massive object similarly bends the light. In this case, the observer is not able to see the objects behind the lens without difficulties. Instead, the object triggering the microlensing would appear with increased luminosity for a period between seconds and years, depending on the duration of the alignment **(21)**.

**3. Target State-of-the-Art**

This section is largely based on the exemplary research brought on in **(22) (23) (24) (25)** and here in short reported.

As introduced before, an automated real-time approach in detecting and classifying transients is highly desirable, given the rate at which the detections per night are raising.

A fully automated real-time classification of those events is desirable for two reasons: first, the need of collecting and processing a huge amount of data; second, the need to discerning what events deserve a proper follow-up for further characterization respect the ones that do not.

A second point derives from the fact that all transients changing their magnitude might represent a wide class of phenomena, the reason why it is important to automatically decide what events are worthy of further investigation (which implies time and costs allocation). Another important problem is represented by the typology of the data, sometimes heterogeneous and incomplete and usually evolving in time **(26)**.

Most of the characterizing information is derived from archival resources and contextualization of other features of the candidate transient (such as its location in the host galaxy). In this case becomes vital that the automated pipeline can discern between noise and artifacts and real events, so as not to miss any important candidate. Such pipeline must have a low rate of false alarm and high completeness, also taking advantage of archives with data from previous (known) events **(27).**

*3.1 Data Reduction Pipeline: key features*

An ideal system should be able to respect severe requirements, listed hereafter and based on **(26)**.

**Automation.** The system must be able to process Gigabytes of data every night and yet allow minimal human intervention. The concept of automated classification and decision-making is probably one of the most representative of this topic since the Authors repeatedly insist on it.

**Real-Time.** Since transients are usually characterized by a very short duration, a human "manual" interevent might not be fast enough to ensure proper classification and prioritization, as well as the activation of a proper follow-up for an event initially unknown (of which we also do not know the duration, until it runs out). Real-time action is, of course, an ideal concept but the practical target is to close the distance between real and ideal case as much as possible until we reach a state of the art in which the system is capable of reacting in a matter of fractions of seconds.

**Reliability.** The more a system is a complex the higher is the probability of errors and malfunctions mostly determined by the failure of its weakest element. Such eventuality must be mitigated at maximum to develop a system with a low failure rate.

**Robustness.** Transients belong to a world of the unknown. Inputs might be very different from expected ones and the system must be able to cope with situations that deviate even much from accounted ones.

**Low rate of false positives.** Incoming data are known to be sparse and heterogeneous. Yet, detections are bound to be contaminated by artifacts and/or already known sources. The system must apply the first filter to discard unwanted inputs and reduce false positives as much as possible.

**Learning.** A system able to act in real-time should be also able to constantly improve its performance by learning over time. This goal can be achieved by creating a proper archive with data from past surveys and other ongoing projects.

**Follow-up.** After the event is detected, little of it is probably known, except basic information such as the Light Curve. In this phase, providing that the event is considered interesting enough, the follow-up is supposed to take over and the system would trigger it in full autonomy. The follow-up (photometric, spectroscopic, multiwavelength) is intended to enrich what is known of the source and by observing key features.

*3.2 Steps in the Classification Process*

Generally speaking, the classification process can be summarized in the following steps **(22)**:

1) Obtain contextual information from a pre-existing archive (point introduced before) and make a comparison with the measured data for the given transient candidate.

2) Asses what are the probabilities that the event belongs to a given class of transients.

3) Run follow-up observations for further characterization to better help a proper classification.

4) Store the data obtained from the follow-up in the archive for the next detection.

In **(26)**, an ideal data reduction pipeline is well shown and explained. The telescopes are represented as a source of incoming heterogeneous and sparse data, along with archival and contextual information (one above all, the spatial location of a given event). As mentioned before, these two inputs are directed to an intermediary Event Portfolio which modifies, time by time, all the data regarding a particular class of transients. After this first step is the Event Classification Engine which compares the fresh data with inputs from a library that gives back a set of probabilities that a given feature manifests, provided that the event belongs to a given class of transients. The final result obtained from the classification engine is a set of evolving probabilities of the candidate transient belonging to various classes. The treated data are now ready to be fed in the Follow-up Prioritization and Decision Engine which will assign a priority level to every possible follow-up measurement among all the possible ones, on account of a cost function. The new data are finally fed back to the Event Portfolio where a new comparison with new fresh data will be made.

The physiognomy here showed is thus of a system capable of learning, survey by survey, how to better discern a class of events from another.

*3.3 Bayesian Network and Light Curve based approach*

Given the large amount of data it is supposed to deal with, its heterogeneity and sparsity, methods based on a Bayesian Network appear to be the perfect candidate to solve the problem of Event Classification, as reported by **(27)** and **(26)** which this paragraph is mostly based on and whose reading is strongly recommended for further deepening. Points derived from different sources will be accompanied by a proper citation.

In order to assign a probability to an event belonging to a given class of transient, it is necessary to exploit all the data gained from the observation, be them fluxes measured at different wavelengths, light curves, etc.

The authors present an interesting demonstration carried out at the Catalina Real-time Survey in which the priors (related to possible classes which a candidate can belong to) are referred to six Cataclysmic Variables (CV), Supernovae, Blazars, Active Galactic Nuclei (AGN), UV Ceti stars and a final class of miscellaneous object among the first five named "Rest". They point out that Light Curves might be a good enough source of data in case they are related to classifiable objects (ex. SNe). In this sense, they report that the Gaussian Process Regression **(23),** used for an automatic classification algorithm, is a valuable tool to gain useful information from the LC of the candidate. In particular, data of interest to be derived are Galactic Latitude, the color of the object in the r-i band, proximity to other objects, and so on.

Another possibility is presented as an LC-based approach. Mahabal and colleagues propose to collect LCs for several objects belonging to a known class and building a probability density function (PDF) to represent its probabilistic structure. This way not only allows to compare LCs from new events with the available PDF but also to enrich the existing archive over several observations of several events.

*3.4 Contextual information and Classifiers*

Putting the data from the first detection aside, the context, as mentioned in the first paragraphs, is also paramount. As introduced before, the light curve of a certain shape might be consistent with several different events, such as Supernovae o variable stars but, as a matter of example, the presence of a galaxy in the proximity of an event makes a Supernova scenario becomes more plausible.

In this sense, it is proposed the use of classifiers to be deployed along a hierarchical path. Some of those classifiers are liable to be more effective if used for a certain event rather than others and key features are used as "filters" in order to exclude that a candidate belongs to a certain class of transients.

An event is filtered through multiple classifiers step by step (from here the definition of "hierarchical" approach).

To better understand the purpose of classifiers, Mahabal et. al. propose a Supernova event as an example. An exploding star can undergo this process only once in its life and a transient candidate with a light curve showing a previous activity before the "main" event could indeed not be a Supernova. Different bits of information called into question different classifiers and even more, than one classifier is liable to be used in the same filtering step **(24)**. The possibility of combining more than one classifier at the same time is presented as a non-trivial possibility, however still under development.

*3.5 Follow-up*

Providing that preliminary observation was made, and that available time and costs are finite, it comes to the issue of identifying the fittest follow-up for the given event. Time, costs, and scientific return are indeed important, but follow-up should also be chosen considering the gain in classification accuracy that a given event could deliver to the system itself.

One way to "set a guideline" in this sense is proposed by **(28)**, who considers the Shannon Entropy as a judgment parameter. The entropy drop related to each possible follow-up would be calculated and an automatic request for the most appropriate choice would be sent to the operator of the available telescopes (be it robotic or human).

Another approach involves human action. The algorithm could be instructed to simply display a list of possible follow-ups to an astronomer, instead of automatically ranking them. The astronomer itself would supply feedback about the most interesting or feasible options so that the algorithm obtains useful information to learn to choose on its own. Mahabal and colleagues specify that this option could be modeled with Multi-Armed Bandit algorithms (MAB), devised by the American mathematician Herbert Robbins in 1952 **(25)**, which can be represented with a slot machine with N levers each characterized by a different return (unknown to the operator). Both are the applicable solutions: exploit the lever that seems to give the highest return; explore different combinations in order to gather information about any possible scheme, initially sacrificing the gain. In MAB the necessity is to find the balance between exploration and exploitation, i.e., maximize the return while minimizing the losses. Practically projecting this example, it is claimed that having enough available telescopes MAB would be used to improve the classification, also improving the background of data working as an archive **(27)**. The return would be represented by the capability of the telescope to assess the class to which any candidate belongs to.

## 4. Catalina Real-Time Sky Survey (CRTS)

The Catalina Real-Time Sky Survey (CRTS) **(29)** is a synoptic sky survey that takes advantage of 3 widefield telescopes: the 0,68 m Catalina Schmidt on Mount Bigelow (Arizona), the 0,5 m Uppsala Schmidt (Siding Spring Survey) in New South Wales (Australia) and the 1,5 m Mt. Lemmon Survey. The CRTS can cover 30.000 $deg^2$ and obtain images of objects down to 21,5 mag over 23 days/lunation, but only at galactic latitudes above 10° not to generate confusion given by objects in the galactic plane, especially in the bulge **(30)**.

### 4.1 Data Reduction Pipeline

The Data Reduction Pipeline exploited by CRTS is in truth largely based on a previous version run in the Palomar-Quest survey.

The approach of CRTS to data reduction and distribution is represented by an open data philosophy since all the transients (and related data like images, LCs, etc) are published within minutes **(31)**. These data can be found available at http://www.skyalert.org/ and, representing the first testbed for a practice that will be brought on in future surveys. This choice implies that external teams are encouraged to further study the gathered data so that the follow-up is largely left to human operators. As expectable, the volume of follow-ups activated is only a small part concerning the number of transient candidates detected **(31)** and this problem is bound to become more insistent as new surveys will come. However, it is highlighted that the data streams from CRTS were, on the other hand, actively used as a test to experiment with new solutions toward an automated data reduction pipeline.

Describing the data reduction more deeply can be firstly outlined the use of the SExtractor photometry software, by the Catalina Sky Survey Telescopes (CSS) in Arizona **(32) (33)**. Transients are identified by comparing the fresh images with source catalogs in order to exclude artifacts and thus false positives. The objects taken from the catalogs are co-added images resulting from the median combination of not less than 20 other images.

Another way the CRTS adopts to find transients is the image subtractions which consists of matching new observations with a high signal-to-noise ratio reference image and subtracting them and the utility of which is highlighted especially for dense stellar fields, i.e., in presence of significant fluxes from mixed sources **(30)**.

As mentioned at the beginning of this paragraph, in order to approach the real-time response, all the candidate transients are processed in loco and sent to the Virtual Observatory (VO) website VOEvent.net, now IVOA.net, and associated with an alert within 5 minutes from the 4$^{th}$ image in a sequence is obtained. It is showed that ~1/200.000 sources are selected as a candidate transient and about 50% of them are then assessed to be a real event, in a process ranging from minutes to hours.

## 5. Transiting Exoplanet Survey Satellite (TESS)

Launched in April 2018, atop of a Falcon 9, TESS is a space telescope devised to reveal the presence of exoplanets by the transit method. According to **(34)** its primary duty is to explore a sky area 400 times greater than KEPLER's, monitoring about 200.000 main sequence dwarf stars, waiting to detect drops in luminosity.

TESS will observe the southern and northern hemispheres 1 year each, dividing each of them into 13 sectors partially overlapped to ensure the presence of constantly monitored regions at the ecliptic poles (Continuous Viewing Zone or CVZ).

These 26 sectors will allow TESS to keep an eye on about 90% of the sky with a temporal cadence of 2 min respect 29,4 min in the case of Kepler **(35)**.

*5.1 Data Reduction Pipeline: Overview*

This paragraph is entirely based on **(36)**, which provides a perfect overview, perfectly in line with the goal of this paper.

It should be noted that the pipeline processes not only scientific data but also engineering data related to the spacecraft itself and only the functions related to the first ones will be considered here.

TESS' Data reduction pipeline is currently in development at the **Science Processing Operations Center (SPOC)** and largely based on software previously devised for Kepler which included pixel-level calibration, background subtraction, aperture photometry, and identification and removal of systematic errors. All the data products generated by SPOC are archived in the **Mikulski Archive for Space Telescopes (MAST) (37).**

Another facility, the **Payload Operations Center (POC)** at MIT, gains all the raw science data via Deep Space Network, sending them to the SPOC which has the goal to analyze light curves searching for transiting exoplanets and assess the probability of whether a given candidate is likely to be a real planet or false detection.

POC and SPOC form together with the TESS **Science Operation Center (SOC)**.

Pixels, light curves, and transit search data are sent back to the POC which transmits them to the MAST and the **TESS Science Office (TSO)** (a catalog with various celestial parameters of up to $10^{10}$ stars).

*5.2 SPOC: The Science Analysis Pipeline*

Based again on **(36)**, another element of interest is the SPOC and among all the components which it is composed of we will focus on the Science Analysis Pipeline which builds diagnostics with the end of ranking in order of priority the candidates for follow-up observation.

SPOC's main components are briefly summarized hereafter in the same order they work.

**Calibration (CAL).** Its main function is to remove artifacts and effects owed to the instrument themselves. A traditional CCD data reduction is performed along with pixel-level calibration and correction owed to the absence of a camera shutter.

**Compute Optimal Apertures (COA).** Here photometric measurements are gained from selected pixels, about each target star passing through the SPOC pipeline.

**Photometric Analysis (PA).** Each image in each frame of the given target star is analyzed and the brightness is measured, also removing flux owed to stars in the background, cosmic rays and performing measures the photocenter of each target star frame.

**Transiting Planet Search (TPS).** Here signatures of transiting planets are detected by putting together the light curves of stars observed over a lunation and in consecutive sectors.

**Data Validation (DV).** When the TPS identifies a planet signature, the DV runs a series of diagnostics to either consolidate or not the confidence that the observed transit candidate is after all a planet. The DV also orders the TPS to run another search for further light curves, in search of pieces of evidence for other planets, running a loop until the TPS cannot find any other signature.

**6. Antarctic Survey Telescopes (AST3)**

The Antarctic Survey Telescopes is a project born from the collaboration between Texas A&M University and the Beijing Astronomical Observatory. Three twin 50cm telescopes (first of which built-in 2012) at the Antarctic Kunlun Station, near Dome A in Antarctica (80°25′S - 77°07′E) **(38)**.

Those telescopes are intended to be fully robotic installation, born to enable efficient sky surveys and to provide a fast response in case of transient detection. The reason for such effort is mainly owed to the fact that the base can spend long periods without human presence and the telescopes must be able to conduct automatic surveys.

Instead of human operators, the main operator of the telescopes is a software called *ast2suite*, developed by **(39)** which orchestrates every aspect of the survey, from telescope pointing to data reduction and archiving.

A strength of the project consists in taking advantage of prolonged periods of dark, during which it is possible to exploit at maximum telescopes with a contained aperture. During polar nights it is possible to set 3-4 months long observations with a cadence that can range from seconds to months.

Information about a data reduction pipeline for exoplanets detection can be found in **(40)** but this case was chosen not to be reported here since it was considered too specific and not applicable to a paper treating all types of transients.

*6.1 Scheduler*

The Authors of this paper wants to specify that this paragraph is entirely based on **(41)** which represents the only source regarding this aspect.

To efficiently plan a sky survey the scheduler must take into account several parameters, such as airmass, observing cadence, survey area, etc, making decisions depending on the observing requirements. For this reason, three different modes (Supernova, Exoplanet, and Special survey mode), each one with its own set of scientific requirements, are created. It is worth mentioning that the Special Mode is intended for those cases in which fast response is needed, i.e. follow-up of Gamma-Ray Bursts or Supernovae. When the mode is triggered by an alarm, the scheduler records the position being observed until that moment, putting that survey in stand by, and quickly points to the "special" event source. Once all special targets are observed, the scheduler returns to the previous survey.

The scheduler works together with the survey system to give a new target each time there is the necessity of pointing a new field. This means that the scheduler is fully autonomous in deciding which target will be the next when to put the survey on standby or operate calibrations. The special mode is the one with the highest priority and the related files ("special files") will be the first to be checked. A proper list is dedicated to these special files and only in case, this is empty the system would switch to the "standard" survey mode. In case the list is not empty the system checks what targets are available to carry out observations, giving higher priority to the nearest target.

As expectable, the telescope spends most of its time in Survey Mode. Similarly, concerning the special mode, the system checks a survey list that contains information about all the fields under control. Fields with the highest priority (according to pre-set criteria) are checked first and the priority level is possibly changed. Information on the observed field is recorded (exposure time, airmass, phase of the moon, coordinates in the equatorial system, etc.) and the survey system goes on stand-by, waiting for a new target or call.

# 7. Large Synoptic Survey Telescope (LSST)

The Large Synoptic Survey Telescope is a facility currently under construction, characterized by an 8,4m primary mirror and located on the El Peñón peak of Cerro Pachón, in northern Chile **(42)**. It will have the capability of overlapping images for a total of 20.000 deg$^2$ in six optical bands (in the range of 320-1050 nm) with an effective system Etendue (also called "Optical Extent" or "Effective System Throughput") of 300 m$^2$ deg$^2$ which is one order of magnitude more than any other existing observatory. Each sky location is estimated to be visited 100 times/year with a 30 sec exposure for each observation **(43)**.

Among the main science, goals are the exploration of the Transient Optical Sky, with an expected detection of several types of transients, among SNe, GRBs, black hole binaries, etc, also paving the path in the detection of a whole new type of transients such as binary mergers and stellar disruption owed to black holes. Microlensing events are also expected to be detected in large quantity especially within the Local Group **(44)**.

## 7.1 Data Products

Regarding the LSST, the very first point worth presenting is represented by the three categories of data products **(45)**.

**Level 1**. Data generated and published continuously every night (within 60 sec from observation), including alerts of an object which brightness or position changed.

**Level 2**. Level 1 data are reprocessed every year for photometric and astrometric calibration and released with full characterization of objects of interest (fluxes, shapes, orbital parameters, light curves, images, list of detected objects, etc.), creating a Dara Release. These data are planned to be stored for the whole lifetime of the LSST.

**Level 3**. LSST's Data management system will dedicate 10% of its capability to user-dedicated processing and storage. This is will allow science teams to use the database infrastructure and store their results inside of it. Proper software will be also put at disposal to facilitate the creation of level 3 data, taking advantage of more than 15 years of efforts put on the LSST.

All Level 1 data and 50% of Level 2 (data release) processing will take place at the Archive Facility at the National Center for Supercomputing Applications (NCSA) in Champaign, which will also serve as a data access center for the US community. The remaining 50% of data processing will be left to the satellite centre (Centre de Calcul de l'Institut National de Physique Nucléaire et de Physique des Particules) in Lyon.

The Base Facility in La Serena has instead the task of serving as a retransmission hub for data uploads to North America and data access centre for the Chilean community.

## 7.2 Data Reduction Pipeline

LSST Pipeline can be represented as a dense group of small pipelines working all together to form a large-scale Data Management System. The available material reports a full characterization of the pipeline which will be not thoroughly reported here, for the sake of brevity. However, an exhaustive description can be found in the second chapter of the LSST Science Book **(46)**.

At the very first level is the **Alert Production**. Astronomers need to search for objects whose flux clearly changes over time and this should be ideally made in the shortest possible time, since field image acquisition. Level 1 data find their place in these first steps where alerts are triggered by the output data stream exiting the Camera Science Acquisition System (SDS) during observations. Images are sent to the **Archive Center** and examined in search of transients within 60 seconds from the shutter closure, by image subtraction (based on **(47)**). Authors point out how the community manifested a strong interest in avoiding filtering alerts before the public distribution so that human operators can fully sift them.

As introduced before, every year a Data Release is organized for the community. The Data Release Pipeline works to generate highly analyzed data products, especially in the case of very faint objects, also covering long time scales. Every year each new run will process the entire survey data set, improving the completeness of the available data. To make a comparison, night (real-time) pipelines are instead based on image subtraction, which highlights the differences between two exposures of the same field and are designed to rapidly detect interesting transient events in the image stream and send out alerts to the community within 60 seconds of completing the image readout, as seen for the Alert Production Pipeline **(48)**.

The LSST is expected to produce several hundreds of Petabytes after the 11$^{th}$ planned Data Release. Authors highlight that it will be an important difficulty to overcome processing such amount of data to transform raw imaged into highly valuable data, also implementing automated data quality assessment and automated discovery of moving or transient sources and making it possible to archive all these data for the community.

## 8. Panoramic Survey Telescope and Rapid Response System (Pan-STARRS)

Pan-STARRS is a survey currently operated by the Institute for Astronomy at the University of Hawaii that runs through two twin telescopes, PS1 and PS2, first of which provides the quasi-totality of the available data and will thus be the main object of the section dedicated to Pan-STARRS.

All the relevant information about the Pan-STARRS survey is easily available at the dedicated homepage **(49)**. PS1 is a 1,8m Ritchey-Chrétien telescope with a 7 deg$^2$ Field of View located in Maui, Hawaii. It is equipped with the largest digital camera ever built, capable of recording 1.4 billion pixels/image and the focal plane is composed of 60 packed CCDs arranged in an $8 \times 8$ array. More information is available thanks to Denneau et.al. which reports that each image is taken with an exposure time of 30-60 sec (enough to see objects down to 22 mag), requiring ~ 2 Gigabytes of storage. Each night the telescope can observe 6000 deg$^2$ of the 30.000 visible from Hawaii, meaning that the entire visible sky can be observed in a matter of 40 hours **(50)**.

It is worth noting that, as well as TESS, the Pan-STARRS takes advantage of the Mikulski Archive for Space Telescope (MAST) of the Space Telescope Science Institute (STScI), to ensure a broad public diffusion of science data **(51)**.

Among the objectives are a series of studies in the Time-Domain of astrophysics including explosive transients, the search of exoplanets, and surveys for microlensing events in the Andromeda Galaxy.

*8.1 Data Reduction Pipeline*

The following paragraph is based on an article by Chambers et.al **(52)**, which reports with great precision how the Pan-STARRS Pipeline is composed and works. The paper is the only article that describes the Data Pipeline and, albeit fully exhaustive, the amount of material would make this article too long. It was thus decided to only introduce all the main components with few lines each to describe their purpose.

**Summit Processing**. Camera and observatory systems run data analysis necessary to support ongoing observations.

**Image Processing Pipeline (IPP).** In this subsystem, raw pixels are processed to obtain calibrated measurements of objects in an internal databasing system. Each image is processed in 30-60 seconds.

**Moving Object Processing System (MOPS).** Detections from inside the Solar System are linked together, and related orbits are determined.

**Published Science Products Subsystem (PSPS).** IPP and MOPS that calibrate measurements are sent to PSPS which creates a high-availability database for the community.

**Institute of Astronomy in Maui (IfA)** coordinates the steps above. In the related paper it is possible to find a full panoramic of the complex system behind data reduction for Pan-STARRS, with a particular emphasis on analysis, calibration and database ingest stages.

From a more dynamic point of view, two responses can be deployed by the system during nightly science operations:

1) Rapid Detection of transient sources, in order to allow a follow-up with other telescopes.
2) Regular Analysis with the purpose of monitoring data quality and for use in longer science projects.

Each image is passed through a processing line to correct instrumental signatures and mainly detect the event sources, by the block CHIP. Astrometric and photometric calibrations are executed by CAMERA. Finally, images are transformed into pixels by WARP. Images of given fields are stacked together, and different images are generated for the nightly stack images or individual warp images. If we are in the second case, warp images can be differenced against another warp of the same night or a reference stack from the given part of the sky.

Pan-STARRS also performed several large-scale reprocessing of data for completed surveys to perform a more detailed photometric analysis on the stack, including morphological analysis appropriate to galaxies.

*8.2 Post Processing*

A brief overview of the post-processing phase seems to be appropriate as well and again **(52)** represents an exhaustive source.

Inside the IPP sub-pipeline is an internal database system called "Desktop Virtual Observatory" or DVO which purpose is to associate multiple detections of the same object, within the context of the photometric and astrometric calibration process.

DVO refers to three main parameters which are the average properties of astronomical objects, the "raw measurements" of the same objects (from which the averages are obtained), and the properties of the images from which the measurement comes from.

To represent the simplest way the DVO works it is possible to imagine a collection of measurements for detections from a set of images loaded inside the DVO itself as well as the metadata describing the images (airmass, exposure time, etc.). The DVO builds astronomical objects based on the uploaded detections with a final goal of creating a database where images and measurements are in relation to one another according to a "one-to-many" relationship and where the same measurements and the derived astronomical objects are related according to a "one-to-many" relationship.

**9. Spitzer Space Telescope**

The Spitzer Space Telescope, named after Leyman Spitzer, was launched in 2003 with a planned mission of 2,5 years.

The **(53)** holds several interesting information about the telescope, starting with the primary 85 cm mirror, which allows the telescope to detect events in the infrared band. Three are the main instruments:

- **IRAC (InfraRed Array Camera)**. It is an IR camera (256×256 pixel) useful to obtain photometric measurements in 4 bands around the infrared (3,6, 5,8, 4,5, and 8,0 microns).

- **IRS (InfraRed Spectrograph).** It is a spectrograph able to work in medium or low spectral resolution between 5,2 and 38 microns.

- **MIPS (Multiband Imaging Photometer for Spitzer).** It is a photometer capable of obtaining images and photometric measurements in 3 bands of the medium and far-infrared (24, 70, and 160 µm).

According to its Handbook **(54)**, Spitzer is mostly committed to executing observations on behalf of the institutes that led its construction, but it also possible for the scientific community to propose a particular survey. The general purposes of the mission are the study of planetology, stellar formation, interstellar medium, Milky Way, and other galaxies, not to mention the study of astronomical transient events.

*9.1 Data Reduction Pipeline*

Plenty of information about the Data Reduction Pipeline of Spitzer is provided by Kasliwal et.al. **(55)** and it will be presented with the study of an unusual class of transients, named SPRITEs, within the SPIRITS (Spitzer InfraRed Intensive Transients Survey).

Such events have no optical counterpart and show an IR luminosity between novae and supernovae with absolute magnitudes ranging from -11 and -14.

The general profile for SPRITEs can be defined accordingly to the following features observed in 14 events with no optical counterpart.

The major components of the data reduction pipeline are presented here below:

**Image Subtraction**. Kasliwal does not lack to highlight that the image differencing code here used was originally developed for the Palomar Transient Factory with few changes which include: the ability to work on co-adds of individual IRAC exposures; masking of regions where the depths are <5, to avoid cosmic rays and detector glitches; the use of SExtractor (valuable tool already used in many other surveys) to select transient candidates from difference images; omission of dynamic photometric-gain matching between reference and science images co-adds.

Reference images come from the archival data, including Super Mosaics (Spitzer Enhanced Imaging Products through the NASA/IPAC Infrared Science Archive) or by stacking prior images in the archive.

Unfortunately, Spitzer difference imaging is subject to a high rate of false positives and the candidate needs to be detected at least twice (in different filters or epochs) to rule out a false positive visually verifying.

**Forced Photometry.** Having made difference images available, a transient candidate is identified assuming a zero flux in the reference image and executing forced aperture photometry on the subtracted image. The sky background is measured within a circular halo (8-16 pixels) around the source and subtracted from the total flux.

**Database and Dynamic Web Portal.** As well as other surveys, data are passed through a web portal and released within few days.

Despite the capability of the pipeline to select candidates on its own, the action of human operators is still needed, and different galaxies are assigned to different team members so that the number of candidates can be restricted. Within one day from the release, team members are assigned with a galaxy and the task to flag interesting candidates. The human action brings to have 1/100 of the objects initially selected by the pipeline itself. After one last step in which contextual information is considered, interesting transients are then announced by Astronomers Telegrams.

## 10. Zwicky Transient Facility

Mount Palomar hosts one of the most famous and productive observatories ever created and its mention seemed a must for this paper. In the following, it is presented the latest survey brought on in this facility: the Zwicky Transient Facility (ZTF).

This survey started in 2017 as the successor of the Intermediate Palomar Transient Factory takes advantage of the 48-inch Palomar Schmidt telescope.

Despite using the same telescope, the ZTF achieved a more than one order of magnitude better volumetric survey speed (spatial volume in which it can detect a transient of given magnitude divided the exposure time) respect the Palomar Transient Factory **(56)**.

*10.1 Observing Strategy and Robotic Observing Software*

This paragraph is entirely described from another publication by Bellm **(57)**, who is currently working as Survey Scientist at the ZTF.

The observing strategy is divided into three main programs which are a public survey, ZTF collaboration surveys, and CalTech surveys which require respectively 40, 40, and 20% of the surveying time. It is also possible that Targets of Opportunity are observed as a response to external triggers. During the public survey mode, two are the main surveys: a Northern Sky Survey during which the observations are focused, with a 3-days cadence, on all the fields whose center is north of -31° of declination; a Galactic Plane Survey which is focused on all the visible fields comprised between -7° and 7° of galactic latitude. Regardless of the survey being run, each field is observed twice with a 30 minutes pause, respectively in g-band and r-band. The totality of the surveys is managed by the Robotic Observing Software (ROS), which is capable of automatically running surveys for long periods. Human operators remain able to control the system (also keep tracking of its performance) and modify some parameters, when necessary. Data (images, catalogs, light curves, etc.) are sent to the IPAC by microwave link, but the system is provided with its 2-weeks worth storage, in case of need.

*10.2 Data Reduction Pipeline*

The general requirement in the conception of the Data Reduction Pipeline was the necessity of a quasi-real-time response (common goal of many other surveys, after all) of no more than 20 minutes between the data transfer to IPAC and its processing. It should be noted that the main pipelines are 9 but, for the sake of brevity, only a brief will be proposed. The material to thoroughly describe the main points of the Data Pipeline have been crossed between the paper previously cited and **(58)**.

An alert system also provides human operators with several contextual information about a detected event (along with the basic measurements, of course) in order to assess whether the candidate is astrophysical or not. This contextual information includes a score from the Real-Bogus machine learning algorithm, a light curve of previous detections, and cross-matches with the Pan-STARRS1 catalog.

In general, the information about such candidates is obtained from positive or negative images (output of the image differencing) and associated with a flux-transient or recurring flux-variable, or a moving object, which represents a set of potential triggers for the alarm. These events pass a first light filter to eliminate any clear false positive, but the rest is left to human judgment.

## 11. Colorado Ultraviolet Transit Experiment (CUTE)

The role of CubeSats in Time-Domain Astronomy is becoming more and more present over the years. One of the most evident pros is given by the fact that small CubeSats can focus on a single target for long periods, despite bigger multi-purpose satellites, such as the Hubble Space Telescope, usually divided between different surveys and teams.

Despite a few projects already operational, a thorough analysis would require a proper space that cannot be dedicated here. For this reason, only one project of particular interest will be presented hereafter, leaving the rest to another paper to be drafted in future work.

The chosen project is called the Colorado Ultraviolet Transit Experiment (CUTE) which is a 4-year NASA-funded project which is led by the University of Colorado, Boulder. The chosen spacecraft is a 6U CubeSat (6 Units = 30cm x 20cm x 10cm) with an expected lifespan of 1 year after being launched in 2020 **(60)**.

### 11.1 Science Goals

CUTE will perform investigations of multiple Hot Jupiters in the Near-UV band ranging from 255 to 330 nm, because of the low optical depth of the escaping gas in the planetary upper atmosphere that can be best studied at ultraviolet wavelengths. The main task is to provide important observations about the atmosphere loss by transit method which allows to "see" the upper layers of an exoplanet's atmosphere that can be inflated up to 3 planet radii. The expected duration is 7 months but, depending on the health state of the instruments, it will be liable for further extension. The exoplanets will be observed around 12 target stars with 10 transits expect for each system **(61)**.

### 11.2 Data Reduction Pipeline

Since many parts are yet to be completed, the data reduction pipeline is only briefly presented in (62) and will be here in short summarized.

The main tasks of the data reduction pipeline will be dark and bias subtraction, corrections for bad/hot pixels, cosmic-ray correction, flat-field removal, spectral extraction, background subtraction, wavelength calibration, and flux calibration.

It is already assessed that some functions will be achieved directly into orbit, as for the case of master dark and bias frames to be assembled on board. Most of the steps will be however carried out on the ground, such as wavelength and flux calibration.

In general, the data reduction pipeline is expected to be flexible enough to be able to account for the effects of possible inflight complications.

## 12. Sardinia Radio Telescope (SRT)

The Sardinia Radio Telescope is a recent and advanced installation with a fully steerable 64-meter single-dish, located in San Basilio in Sardinia (Italy), inaugurated in 2013.

Its capability allows high efficiencies at frequencies up to 115 GHz, but the most exploited range is usually between 0,3 and 100 GHz **(63)**.

Along with the other installations forming the European Pulsar Timing Array (EPTA), the SRT will provide coverage for pulsars, and yet it investigates several other transients. Among these, it is possible to mention Active Galactic Nuclei and Gamma-Ray Bursts, not to mention the investigation of Gravitational Waves.

An interesting peculiarity is given by the presence of 1116 actuators whose duty is to correct deformations in the main mirror owed to gravity and errors owed to wind-related effect and temperature **(64)**.

*12.1 Data Reduction Pipeline*

First, the SRT takes advantage of software named *ScheduleCreator* whose function is to set a schedule for including all the possible modes the SRT can exploit.

The telescope is mainly directed by the SRT ExpAnded Data Acquisition System (SEADAS) which works for antenna pointing and configuration and data acquisition. The last point, in particular, is managed by communication with other software tools left running in the backend' server **(65)**.

The most recent concept for the SRT pipeline, strictly speaking, can be found in **(66)**. Such an article shows more than one pipeline, the reason why a list will be exposed hereafter.

**Radio Pipeline.** This pipeline will be able to observe pulsars in all the bands the SRT can exploit (P, L, C, K, and S-band) by the intervention of software named *presto* **(67)** which will be aided by a python program developed by the team itself. Being the location characterized by high radio interference, the pipeline will also be optimized for Radio Frequency Interference excision while the machine learning algorithm will take care of an automatic selection of candidates.

**Gamma-Ray Pipeline.** This pipeline has its input from online archives that harvest data from AGILE and FERMI orbiting telescopes. Those data are used to trigger a follow-up of the source and, thanks to a Bayesian and machine learning-based approach, a galactic progenitor will be identified as a candidate source.

**X-ray Pipeline**. Similarly, this pipeline downloads data from Nasa's NuSTAR telescope. It must be highlighted that Imaging X-ray Polarimetry Explorer (IXPE) will take the place of NuSTAR itself, being a very similar project and being the team involved in it.

**13. Gamma- Ray Light Detector (AGILE)**

Launched in 2007, AGILE is an Italian observatory deputed to the observation of high-energy events, being able to detect and image particles between 30 MeV and 50 GeV and 10-40 keV.

Every useful information about AGILE is reported in its Handbook **(68)**. It is mainly addressed to observing GRBs, Active Galactic Nuclei, Galactic Sources, Pulsars, Binary Systems, Supernova Remnants, and others.

Among several instruments, observations will be operated by: Gamma-Ray Imaging Detector (GRID), devised to obtain a large field of view and work in the 30 MeV-50Gev energy range; Super-AGILE detector, intended for observation of sources both in the Gamma and X-ray bands.

*13.1 Data Reduction Pipeline*

The Data Analysis pipeline is reported in the sixth chapter of the handbook.

The chapter shows how science data are sent to the ground facility in Malindi, allowing the production of Level 1 data (science data passed through minor corrections, such as background rejection, attitude, etc.). Level 2 data will be then ready to be treated with proper software, allowing the personnel to operate a full science analysis for each point-like candidate source of interest.

The goals of the data analysis by AGILE can be summarized as: analyzing gamma and x-ray data within 1 hour from detection; making results available for the public via the web; creating alerts of GRB in 2 minutes maximum; allowing other teams to analyze specific gamma-ray sources.

Another important point is the capability of AGILE of observing sources both in the gamma and x-ray band. As a matter of example, the gamma-ray outburst from the binary V404 Cygni **(69)**.

## 14. Robotic Optical Transient Search Experiment (ROTSE)

The ROTSE is a project composed of four telescopes designed to observe the optical afterglow of gamma-ray bursts **(70)**.

After two predecessors, ROTSE-I and ROTSE-II, the ROTSE-III system went online and started working with a CCD camera composed of a 2048 x 2048 pixel matrix. Another interesting point, proposed in **(71)**, is the automated data acquisition software. The ROTSE can work on its own thank to a fully automated system created for ROTSE-I and managed by several daemons (automatic programs working in the background): the clamshell daemon (*clamd*) managing the aperture and closure of the clamshell; the camera daemon (*camerad*) which manages the CCD camera; the weather daemon (*weathd*) which monitors the weather and can command the closing of the clamshell in case of bad conditions; the mount daemon (*schierd*) which manages to point and tracking; the astronomical scheduler daemon (*astrod*) which schedules observations, system startup, and shutdown, also managing the queue and alerts for GRBs; the alert daemon (*alertd*) which manages and triggers alerts about sudden events.

*14.1 Data Reduction Pipeline*

Information about the data reduction is retrieved from **(72)** in chapter 8.

The first concept of a pipeline for ROTSE-III involves the acquisition of hundreds of images per night in a fully automatic regime. The general idea is to create a closed-loop cycle.

When an event is detected, the Camera Server Daemon (*camserverd*), acting as the server on the camera computer itself, records a new image to the archive. Consequently, a script called *sexpacman.pl* wait for new image-related links, and when a new one appears it operates dark and flat fields correction on the given image, also taking advantage of SExtractor to produce a list of objects from the image itself. In the case of a GRB, the image acquires a higher priority and the newest outbursts outrank the older ones.

The list of objects created by SExtractor is then passed to a script called *idlpacman.pro* which reads the first file, calibrating it against the United States Naval Observatory (USNO) catalog to generate a list of R-band magnitudes and locations for the candidate sources. Up to this point, 45 seconds are usually passed.

Advancing with the hypothesis of a GRB, data structures of calibrated lists allow the first identification of few objects. At this point, variable objects easily appear in the images and can be flagged with no further steps. If it is not the case and the candidate is a real transient the image is divided into a grid. Magnitudes related to stars are obtained that's to such grid which allows recognizing an array of magnitude offsets by which all magnitudes are adjusted.

## 15. James Webb Space Telescope (JWST)

The JWST is usually presented as Hubble's successor and, scheduled to be launched in 2021, is a large infrared telescope with a 6,5 m primary mirror. Named after the former Nasa administrator James Webb, the telescope will operate to achieve many goals, transients study included. In particular, it will be able to study exoplanets by transit and even gather information on their atmospheres **(73)**.

*15.1 Data Reduction Pipeline*

Little is found in literature about the future data reduction pipeline of the mission and the main source about the following information is represented by **(74)**.

The JWST pipeline is expected to fully involve the science community, and to improve over time thanks to lessons learned and new best practices. The science community will be put in condition to download large amounts of data from JWST observations and to track the health condition of the instruments over a long period.

Once the spacecraft is into orbit, a series of null-tests will be carried out, also on the proposal of the community itself and with the goal of testing telescope capabilities and to find an agreement on how to get rid of known problems (for example, ramp effect).

A first release called Early Release Observation (ERO) will be made so that the community can be encouraged and involved in working with the telescope, also having the possibility to deal with real data from real transit detections and learn how to work with them.

Another, not yet defined, release (Early Release Science or ERS) will be complementary to the ERO and its main goal will be to allow the community to understand the performances of the telescope before the submission of the first proposals for the mission.

## 16. Very Large Array (VLA)

The Very Large Array is an interferometer completed in 1980 dedicated to a broad range of studies.

Its website **(75)** contains all the relevant information, starting from the main features. The installation is composed of 27 antennas with a diameter of 25 meters, located to form a "Y" which lines are 21 km long. Thanks to the principles of interferometry, the whole complex can act as a giant antenna of 40 km in diameter.

Among the main scientific goals is the study of quasars, pulsars, radio galaxies, black holes, gamma-ray bursts, and transients in general.

*16.1 Data Reduction Pipeline*

For this part, it will be presented an interesting system proposed by Law et. al. in 2018 and called *Realfast* **(76)**.

The main reason that led to the birth of *Realfast* is the volume of data coming from data collection with a millisecond cadence. This novel system would instead operate only during a few moments, when the transient manifest, reducing the volume of data by a factor of 1000 (according to what the Authors claim).

The authors provide a full description of the whole system (algorithm, hardware, software, etc.), but in this paragraph, only the main features of the pipeline are reported. For further details, the reading of the original article is strongly recommended.

The first two great actors of the pipeline are the WIDAR and the Correlator BackEnd (CBE). The first one manages the 1$^{st}$ level correlation and the second one manages the 2$^{nd}$ level, forming together the correlator of the VLA. Both the blocks allow the generation of visibility data (meaningful, structured data).

Hierarchical Decision making allows a reduced data rate yet maintaining an ideal sensitivity and computational efficiency. Constant sources are early subtracted by subtracting early visibility in time on timescales lower than the VLA fringe rate (which results in about 1 s). This way allows to recognize transient candidates by thresholding all the images, also eliminating the need for a source catalogue. Candidates are saved to a database.

All in all, the *Realfast* will be required to generate structured data for few tens of events a day, also expecting some false positives. The "winning" candidates will be directly sent to the National Radio Astronomy Observatory (NRAO) archive.

Transient of interest will be shared with the community and distributes by the FRB VOEvent protocol.

## 17. Wide-Field Infrared Survey Telescope (WFIRST)

The WFIRST is a Hubble-sized space telescope with a mass of 4166 kg and its 300 MPixels Wide Field Instrument (WFI) camera will have a field of view 100 times larger than Hubble's itself.

NASA will implement the WFIRST program on a 2.4-meter AFTA (Astrophysics-Focused Telescope Assets) telescope, donated to NASA by another agency.

To be launched in the mid-2020s, this telescope will be able to conduct surveys in the optical and infrared bands, providing new important hints about three main topics: Supernovae, Exoplanets habitability, and Dark Matter.

A spearhead of the mission will be the Coronagraph Instrument (CGI) which will consist of a technology demonstration for possible future missions aimed at detecting signs of life in the atmospheres of Earth-like exoplanets. It will also be capable of directly imaging planets similar to those in our Solar System, measuring for the first time the photometric properties of the mini-Neptune or super-Earth planets. The instrument will be able to suppress the starlight by a factor of 1 billion, far better than current state-of-the-art ground or space-based capabilities.

All the information above and other relevant features can be retrieved from the WFIRST official Website **(77)**.

*17.1 Data Reduction: PISCES Spectrograph*

The newest concept for a data reduction pipeline is strictly related to a proposed instrument for the CGI: Prototype Imaging Spectrograph for Coronagraphic Exoplanet Studies (PISCES) **(78)**.

This particular instrument was proposed according to the WFIRST CGI requirements and to allow spectroscopy from direct imaging of exoplanets to analyze the atmospheres of Earth-sized rocky planets.

The first step is the calibration process in which centroids of detected images are found. The idea is to create a "global wavelength calibration map" to allow the creation of 3D wavelength-dependent cubes using the detected data. It is highlighted how managing the size of wavelength intervals it is possible to control the precision of the calibration as the maps become more accurate as the intervals become smaller.

The data reduction is widely based on a pipeline created for the Gemini Planet Imager, the main instrument of the Gemini South Telescope (Cerro Pachón, Chile) and called "GPI Pipeline" **(79)**. By choosing different algorithms, among the available ones, it is possible to obtain different variants of the data reduction.

The final performance is represented by a computationally fast spectra extraction and simulations demonstrated high accuracy in the extraction itself.

## 18. EUCLID

Euclid is a mission proposed by the European Space Agency, scheduled for launch in 2022.

Its capability of surveying the visible and near-infrared band will allow the mission to better understand what is behind dark matter and dark energy by measuring the acceleration of the universe (redshift). The mission will also investigate astronomical transients, especially SNe, exoplanets, and gravitational lensing **(80)**.

Euclid will be equipped with a 1.2 m diameter and two main instruments: a photometer in the visible domain (VIS), and a photometer/spectrometer in near-infrared (NISP, 900-2000 nm). The spacecraft will operate in the Lagrange-2 point and the mission will nominally last 6 years. The extragalactic survey will cover 15.000 deg$^2$ and around 15 billion galaxies, the deep survey will cover 40 deg$^2$ (80 times the moon) and about 10.000 galaxies **(81).**

*18.1 Data Reduction Pipeline*

Laureijs et.al. **(82)** provides a first explanation of what concerns the data reduction pipeline. The main body of the pipeline will be the Science Ground Segment (SGS) which works with raw data processing up to the final data products. VIS and NISP data will be merged with the ground-based data to derive interesting surveys whose final purpose will be indeed the accomplishment of mission objectives.

The SGS will be backed by a distributed Euclid Archive System (EAS) and will be requested to work with large amounts of data, also providing a quality check at each step of the processing. Mission management and science operations will be left to ESA, while the Euclid Consortium (which would represent the Principal Investigator) will care about covering science-related components and sharing data products with the scientific community. Another important factor will be represented by the Operational Units (OUs), i.e. groups of scientists who will be in charge of developing data processing algorithms.

As the first step, the Science Operation Center (SOC) in Madrid (ESAC) receives telemetry and various raw data (Level 1 data), sending them to first processing and then storing them in the EAS. Dubath et.al. **(83)** better explore this step, explaining how cleaning and corrections such as bias subtraction, flat fielding, cosmic ray removal are operated by processing functions named VIS, NIR, SIR, and EXT. Such functions take care of visible images, near-infrared images, near-infrared spectra, and frames acquired from the ground, in this order. A further function, the MER, is entitled to merging all these outputs, creating the product for the Euclid Catalogue with source identifications, calibrated flux measurements, and spectra. Authors spend few lines to highlight the possible eventuality that a source can be especially active in one photometric band. In this case, the final source detection can indeed happen only at MER level, during this comparison.

Level 3 data are produced by the other three functions: SPE, PHZ, and SHE. The first two derive the spectroscopic and the photometric redshift measurements, and SHE works in galaxy shape determination. Another function called SIM backs the whole process, by creating simulated data to be compared, in parallel, with real outputs.

*18.2 Joining forces with LSST*

A paper by Rhodes et.al. **(84)** proposes interesting teamwork between the LSST and Euclid. This paper bases the confrontation respect several types of astronomical objects but, to this paper, only the transients-related parts were noteworthy. This topic is indeed interesting but only a few points will be reported here, as an example, and to give the reader a first hint of what the common advantages would be.

Recalling that the LSST provides Time-Domain imaging in the southern hemisphere in six bands (u, g, r, I, z, y), the Authors see the two telescopes as complementary assets, and the cooperation between this ground facility and the Euclid in orbit is seen as a winning choice.

As a matter of example, the LSST would work to confirm transient sources that lay beyond the limits of Euclid, also countering problems given by the presence of bogus and not interesting objects. Forced photometry would also be interesting to be carried out in both surveys, providing that the flux measured by one is obtained thanks to the aperture of a source detected by the other survey.

Another strength of this proposal lies in the Euclid Wide Survey, which can be carried out only with a single visit at the time. For transient detection, this is not the best approach (and yet transients remain objects of interest for Euclid), but Euclid could provide additional data to the LSST survey, such as spectroscopic redshifts (by NISP) for a fraction of Ia-type SNe detected by LSST itself. This would be, for example, a way to improve the elimination of systematic biases when using photometric redshifts alone.

## 19. Discussion

After presenting the desired state of the art and some of the most active and famous surveys, we finally arrive at the necessity of comparing those two realities. Before starting the comparison, the Authors of this paper want to highlight that in many cases all the information necessary to expose a given survey was available only thanks to those who are actively involved in the surveys themselves. For this reason, the typical spirit of an article review will manifest in this section.

For the sake of readability and order, the comparison will be carried out by survey.

**Catalina Real-Time Sky Survey.** The effort in this survey is largely based on previous ones (See Palomar-Quest) and the human component is still predominant, not to mention that the whole processing may take minutes as well as hours. Near Real-Time processing is supposed to take not more than a fistful of minutes. On the other hand, the open data philosophy allows several teams to work in parallel at the same time and guarantee an efficient transient detection over a (potentially) short period. In this sense is clear how automation is given a lower priority.

**Transiting Exoplanet Survey Satellite.** TESS brings KEPLER's performance to a new level and introduces a new dimension of automation in exoplanets survey, being able to orchestrate a complex data pipeline. This is of course made possible by a strong union between in-flight and on-ground operations.

**Antarctic Survey Telescopes.** This is a particularly noteworthy survey, in which the necessity for automation in an isolated, infrequently crowded facility, led to an interesting product. Prolonged periods of dark are exploited at best with a fully automatic telescope and related pipeline, from the survey scheduling up to the final product. Unfortunately, no data were found about the response time in case of a sudden fast event.

**LSST.** The LSST is the spearhead of its kind. Its level 1 data products are generated within 60 seconds from the event itself and in this case, alerts are included as well. The automatic operations carried out by the data pipeline are associated with a further level of analysis that can be carried out by human teams to further filtering or reconsidering the created data products.

Also, yearly data releases will allow recovering missed events, even though no alarm is triggered due to, for example, faint events. This thick net of data processing represents a promising tool for future surveys for which the LSST will be a precursor.

**Pan-STARRS.** This survey is characterized by a highly automatic data pipeline, strongly orientated to public diffusion of the data product. Data reprocessing (already seen in the LSST) is also important for ensuring not to miss any transient candidate not promptly recognised at the right time. It is however clear that the Pan-STARRS still relies on a high presence of man-work, but it is also clear that is capable of keeping up with other similar surveys, with a response time below the minute. The Pan-STARRS is thus a Near Real-Time facility with a right, but a milliseconds-response is still far from the current level.

**Spitzer Space Telescope.** Along with the more famous Hubble Space Telescope, the Compton Gamma-Ray, and Chandra, Spitzer is one of the four Nasa's Great Observatories. An important problem comes to attention: the high rate of false-positive and the necessity of multiple detections of the same object which, of course, affects the real-time response capability of the observatory. It also requires a manual (visual) check of the images to rule out false positives, so that the pipeline still needs a human "step" to complete its data processing and production. Within one day from public data release, human teams are indeed vital to rule out false candidates, leading to an almost 1:100 proportion between automatically detected candidates and confirmed candidates, which highlights the limits of this survey.

**Zwicky Transient Facility.** The presence of the Robotic Observing Software allows long surveys fully conducted in automatic. No human intervention is needed, except for monitoring and parameters modification. Its autonomy in terms of storage is also praiseworthy and allows the facility to function in isolation for up to 2 weeks. Its near-real-time response capability is to be considered one of the highest levels in the currently ongoing surveys.

**Sardinia Radio Telescope.** The existence of different pipelines optimized for specific observations is indeed intriguing. Also, the versatility of the telescope allows it to operate in several fields but, unfortunately, little material was found about any connection with the scientific community.

**AGILE.** A strength of AGILE that stands out is the response of 2 minutes for GRB alerts. The scientific community will benefit from an important service, being able to promptly steer ongoing surveys on the gamma source in a very reduced time. Also, the capability of conducting X and Gamma-Ray surveys at the same time is not a common feature and it represents an important tool to make exciting science.

**ROTSE.** As well as the Antarctic Survey Telescopes, ROTSE combines a fully automatic response to sudden events with fast data processing which happens in a matter of minutes. The inheritance from ROTSE-I helped to create a good example of fast-response transient detection and classification survey.

**James Webb Space Telescope.** As Hubble's successor, James Webb will be almost fully dedicated to the scientific community. Several teams will be able to benefit from its great capabilities and a large number of papers and contributions, in general, is bound to see the light from the early years. The study of rocky exoplanet atmospheres will probably be one of the most important contributions of the telescope. It would have been interesting to characterize the expected response time of the telescope in case of fast transient, but no information was found about this point.

**Very Large Array.** Despite its longevity, the VLA is still able to deliver valuable science to the community. In particular, the *Realfast* pipeline would allow fast response and a fast data-sharing with the community, as well as other younger surveys.

**WFIRST and EUCLID.** Those two telescopes have yet to demonstrate their capabilities on a mission and to little was found about their data processing.

WFIRST showed a preliminary high accuracy in creating final data products and this will probably ensure a low rate of false-positive without any further human intervention.

EUCLID, on the other hand, shows an interesting concept in which characterized by parallel functions working at the same time on different features of the acquired raw data. This will probably ensure fast processing and data production for the scientific community, not to mention the use of simulated data to constantly provide a comparison in the processing.

Its potential coupling with the LSST seems also to be the perfect receipt for a prolific science program.

## 20. Conclusions

In this paper, a review mapping of the state-of-the-art of Time-Domain Astronomy Surveys (both on the ground and in orbit) is proposed. Astronomical Transient Events were introduced, listing some of the most representative and well-known types. Based on the literature available, it was also showed what are the desirable features of the ideal real-time detection and classification system and those that could be the most indicated approaches to the problem.

In the second instance, several surveys were presented to outline the respective data reduction pipelines in their main features.

Finally, a comparison between the presented surveys and the expected level for the state of the art was made to show the gap, if any, between the current level and the desirable one. It is eventually clear that the human component is still dominant in the majority of the pipelines since no system cannot be separated from proper monitoring. On the other hand, several promising surveys (such as the LSST and the Zwicky Transient Facility) show great improvement concerning the previous generation of and it is reasonable to expect even more automated systems for the times to come. Another challenge will be closing the distance between the current near-real-time capability and real-time capability (milliseconds-response).